# Cool Outflows and HI absorbers with SKA

Raffaella Morganti[1,2], Elaine M. Sadler[3], Stephen J. Curran[4] and HI SWG Members

[1] *ASTRON, the Netherlands Institute for Radio Astronomy, PO Box 2, 7990AA Dwingeloo, The Netherlands*
[2] *Kapteyn Astronomical Institute, University of Groningen, P.O. Box 800, 9700 AV Groningen, The Netherlands*
[3] *ARC Centre of Excellence for All-sky Astrophysics (CAASTRO) and SIfA, School of Physics A28, University of Sydney, NSW 2006, Australia*
[4] *School of Chemical and Physical Sciences Victoria University of Wellington, New Zealand*

*E-mail:* morganti@astron.nl  ems@physics.usyd.edu.au

HI 21-cm absorption spectroscopy provides a unique probe of the cold neutral gas in normal and active galaxies from redshift $z > 6$ to the present day. We describe the status of HI absorption studies, the plans for pathfinders/precursors, the expected breakthroughs that will be possible with SKA1, and some limitations set by the current design.

## 1. Introduction

One of the main goals of SKA is to trace the neutral phase of hydrogen, the most abundant element in the Universe, out to high redshift and connect this to the formation and evolution of galaxies across cosmic time. Observing the neutral hydrogen (HI 21cm) line in emission can achieve part of this task, but an important and complementary study uses the HI 21cm line detected in absorption against strong radio continuum sources.

Studies of *associated HI 21cm absorption* (HI located in and around the host galaxy of a radio source) allow us to understand the HI content of individual galaxies (as a function of their morphological properties, redshift and environment), the structure of the central regions and the feeding and feedback of active galactic nuclei (AGN).

Studies of *intervening HI 21cm absorption* (HI absorption toward radio-loud background sources, see e.g. Wolfe & Davis 1979, Kanekar & Briggs 2004), allow us to measure the number density of 21-cm absorbers and constrain the evolution of cold gas in normal galaxies over more than 12 billion years of cosmic time. HI 21cm absorption lines can also be used to probe fundamental constant evolution, by comparing the HI redshifts to those of redshifted ultraviolet lines (e.g. Wolfe et al. 1976; Curran et al. 2004, Kanekar et al. 2010), mm-wave rotational lines (e.g. Drinkwater et al. 1998, Carilli et al. 2000, Murphy et al. 2001), and OH lines (e.g. Darling et al. 2003, Chengalur & Kanekar 2003, Kanekar et al. 2012).

Below we briefly review the status of HI 21cm absorption studies and the most important perspectives that will be opened up by SKA1. HI absorption studies have both advantages and disadvantages compared to HI emission studies. The detectability of an absorption line is essentially independent of redshift, since it depends only on the strength of the background continuum source, so HI absorption measurements can probe the presence of neutral hydrogen in high-redshift galaxies where the HI emission line is far too weak to be detectable. HI 21cm absorption can also be detected and studied at very high spatial resolution (including at milli-arcsecond scales with VLBI if the background source remains bright enough), which is not possible for HI emission studies. Thus associated HI absorption has been used to probe the circumnuclear regions of radio-loud AGN on very small scales. Furthermore, given that the absorbing gas *has to be located in front* of the radio source, the kinematics of the gas can be more easily disentangled, therefore making studies of infall/outflows less affected by ambiguities.

The observed HI 21cm optical depth $\tau$ is related to the HI column density $N_{HI}$ (in the optically-thin regime) by the expression

$$N_{HI} = 1.823 \times 10^{18} \, [Ts/f] \int \tau \, dV \quad (1)$$

where $Ts$ (K) is the harmonic mean spin temperature of the HI gas, $f$ is the covering factor (i.e. the fraction of the background radio source that is covered by intervening gas) and $\Delta V$ the line width in km s$^{-1}$. If the HI column density can be measured independently (e.g. from the optical damped Ly $\alpha$ absorption profile along the same sightline), then the 21cm optical depth provides information on the spin temperature and covering factor of the neutral gas. Conversely, if some reasonable assumptions about the spin temperature and covering factor can be made (e.g. Braun 2012), then an HI column density can be estimated even without optical Ly $\alpha$ measurements. As can be seen from equation (1), at fixed HI column density the observed optical depth rises as the mean spin temperature decreases, so HI 21cm absorption measurements are most sensitive to the cold neutral gas within galaxies (Rao & Briggs 1993) – in contrast to HI 21cm emission-line and Ly$\alpha$ absorption studies, which are sensitive to both warm and cold gas.

One disadvantage of HI 21cm absorption studies is that their view of the HI distribution is limited to the regions where the background continuum is present. To partly compensate for this, it is important to take advantage of other gas diagnostics (i.e. multiwaveband data) and information about other gas phases (ionized and molecular) where possible.

## 2. Associated absorption

### 2.1 Motivation and previous work

HI 21cm absorption in radio AGN has been detected and studied for many years, starting with the detection against the core of the radio source Centaurus A (Roberts 1970) and the nearby spiral galaxy NGC 4945 (Whiteoak & Gardner 1976). This associated HI absorption has been used to trace the gas in central regions of radio AGN, as well as to probe the extreme physical conditions due to the interplay between the energy released by the active black hole (BH) and the ISM.

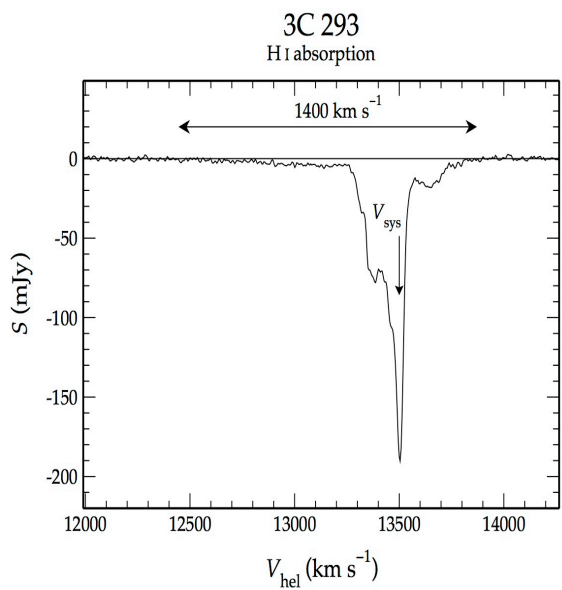

*Figure 1 – HI absorption in 3C293, as detected by the WSRT. A broad, shallow HI absorption is detected as well as a deep, narrower component. The latter is associated with the circumnuclear disk/dust-lane while the former represent gas associated to a fast outflow (~1400 km s$^{-1}$, Morganti et al. 2003). Because of its location (~0.5kpc from the nucleus) the outflow appears to be driven by the radio jet/lobe. This profile shows the need for a stable instrumental bandpass as discussed in Sec.2.*

The detection of HI 21cm absorption in active galaxies has been often considered a **tracer of circumnuclear disks**. Several such cases have been studied, e.g. Beswick et al. 2004, Struve & Conway 2010, 2012, Peck & Taylor 2001, Morganti et al. 2008. Even more intriguing has been the discovery that HI can also be associated with kinematically extreme phenomena. Fast ($\geq$ 1000 km s$^{-1}$) and massive (up to 50 M$\odot$ yr$^{-1}$) **outflows** have been found traced by HI (Morganti et al. 2005, Teng et al. 2013, Morganti et al. 2013 and ref. therein). Gas outflows driven by AGN and supernovae (SNe) are a key ingredient in current models of galaxy evolution (e.g. Croton et al. 2006; Hopkins & Elvis 2010; Schaye et al. 2015), since they can efficiently heat up or even expel gas from a galaxy. This feedback process regulates galaxy growth over cosmic time by stopping or regulating further star-formation and nuclear accretion through feedback processes. Feedback therefore appears to be a fundamental process that affects the bulk of the baryons in the universe, and more detailed observations of gas outflows are needed to build a coherent picture of their role in shaping galaxies over time. High spatial resolution observations (sub-arcsec and VLBI) have shown that outflows can be co-spatial with bright radio components, suggesting the radio jets are the driver of such outflows (Morganti et al. 2013, Tadhunter et al. 2014). The fast outflows have been traced using different phases of the gas (including molecular gas, e.g. Feruglio et al. 2010, Alatalo et al. 2011) showing that they are truly multi-phase structures and that cold gas can live in a harsh environment. HI and molecular gas represent the dominant component of the outflows and, therefore, their detection is highly relevant to quantify the impact of feedback and the relevance of radio plasma in this process. Even low power radio emission can provide the driver for gas outflows (e.g. Nyland et al. 2013, Combes et al. 2013), opening possibilities for many more cases to be discovered by more sensitive radio telescopes and in particular SKA1.

Finally, the gas in the nuclear regions can have an important role triggering the nuclear activity. For example, **infalling** gas, whether from an accretion disk or an advective flow, is an essential ingredient of AGN activity, as it can feed the central engine, turning a dormant supermassive black hole into an active one. Suggestions that this can be traced by HI 21cm absorption have been presented e.g. in van Gorkom et al. (1989) and Morganti et al. (2009). However, the search for infalling gas has been, so far, less successful. This could suggest that the HI does not always provide the fuelling or that the amplitude of the infalling gas is not large enough to be easily isolated from the regularly rotating gas or fast outflowing component.

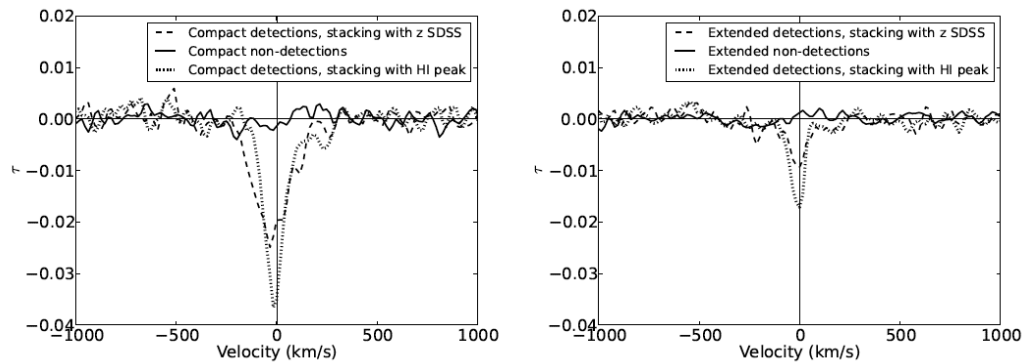

*Figure 2 - Stacked HI profiles for compact (left) and extended (right) sources (from Gereb et al. 2014). The profiles show the higher optical depth and the broader width of the stacked profiles of the former, suggesting differences in the medium in which the two groups of objects are embedded.*

Gereb et al. (2014a,b) present an overview of the HI content, detection rate and kinematics of associated HI 21cm absorption in a large sample of nearby radio galaxies. To date, this work most closely mimics the kind of *"blind survey"* which will be done with SKA pathfinders and precursors and, to even higher sensitivity, with SKA1. The Gereb et al. (2014a,b) work expanded the search for HI 21cm absorption down to 50 mJy radio sources and made use of stacking techniques not previously used for extragalactic absorption studies. This survey found a high detection rate of HI absorption (~ 30 %), which does not appear to be biased towards brighter continuum sources. This is an important result, because it suggests that HI 21cm absorption studies can be extended to even lower radio fluxes, which is important input for future surveys. The stacked profiles of detections and non-detections reveal a clear dichotomy in the presence of HI, with the detections showing an average peak optical depth $\tau = 0.02$ corresponding to an estimated HI column density $N_{HI} \sim 7 \times 10^{20}$ cm$^{-2}$ (for $T_{spin} = 100$ K and covering factor $f = 1$), while the non-detections remain undetected with a peak optical depth upper limit $\tau < 0.002$ corresponding to $N_{HI} < 2.26 \times 10^{19}$ (Ts/100 K) (dV/FWHM km s$^{-1}$) cm$^{-2}$.

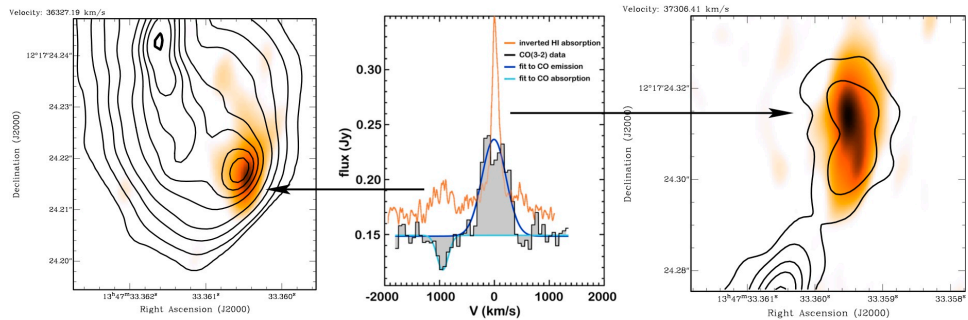

*Figure 3 - Illustrating the importance of locating the absorbing gas, the case of 4C12.50 (Morganti et al. 2013). The central image shows the integrated HI profile (orange) superimposed on a CO emission profile for comparison, while the two outer plots show the location of two HI clouds based on 21cm absorption measurements made at high angular resolution.*

Thus, 30% is a representative detection rate of HI in AGN. Comparing these results with what obtained in the study of HI for nearby early-type galaxies (e.g. Serra et al. 2012), one can conclude that HI 21cm absorption can be used to trace the HI content of these galaxies, and so could be used to trace the presence of gas in high-$z$ objects (instead of HI emission for which there will not be enough sensitivity). Orientation effects due to the distribution of HI in a disk-like structure can be partly responsible for the observed dichotomy in optical depth, although some galaxies must be genuinely depleted of cold gas. Young, compact radio sources show a higher HI detection rate, confirming and expanding the results from smaller samples (e.g. Gupta et al. 2006, Chandola et al. 2011). The HI 21cm absorption lines in these compact sources also show higher optical depth and large FWHM compared to extended sources, strengthening the idea that the compact sources are richer in nuclear gas.

Finally, it is important to emphasise the crucial advantage that high spatial resolution provide for disentangling all these structures. HI 21cm absorption observations have revealed regularly rotating structures in a number of compact sources (Struve & Conway 2010, 2012, Peck & Taylor 2001). However, extremely important is also that VLBI observations have confirmed, in a number of cases, the key role that radio jets are playing in outflow of cold gas by locating (at the end of the radio jet) the fast outflowing gas. The most impressive case is 4C12.50, shown in Fig. 3 (Morganti et al. 2013).

## 2.2 Associated HI 21cm absorption in high-redshift radio sources

Detections of associated HI 21cm absorption in the high redshift Universe remain extremely rare, with only two detections to date at $1 < z < 2$ (Ishwara-Chandra et al. 2003; Curran et al.

2013) and two at z ≥ 2 (Uson et al., 1991; Moore et al., 1999). In contrast, more than 40 associated HI 21cm absorption systems are known at 0.1 < z < 1 (e.g. Carilli et al. 1998; Vermuelen et al. 2003), and the detection rate for compact radio sources is typically ~30% in this redshift range (e.g. Vermeulen et al. 2003; Gereb et al. 2014).

Curran et al. (2008) have proposed that the low detection rate of associated HI absorption at z > 1 is due at least in part to the selection of bright radio sources with known optical redshifts as targets for an HI search, since most of these objects are powerful QSOs with a rest-frame ultraviolet luminosity above $L_{UV} \sim 10^{23}$ W Hz$^{-1}$ (see Figure 4) in which the interstellar gas is likely to be completely ionised (Curran & Whiting 2012). The broad frequency coverage and large instantaneous bandwidth of SKA1 will remove this problem by making it possible to carry out genuinely blind HI 21cm absorption surveys out to z > 6 without the need for optical pre-selection of targets.

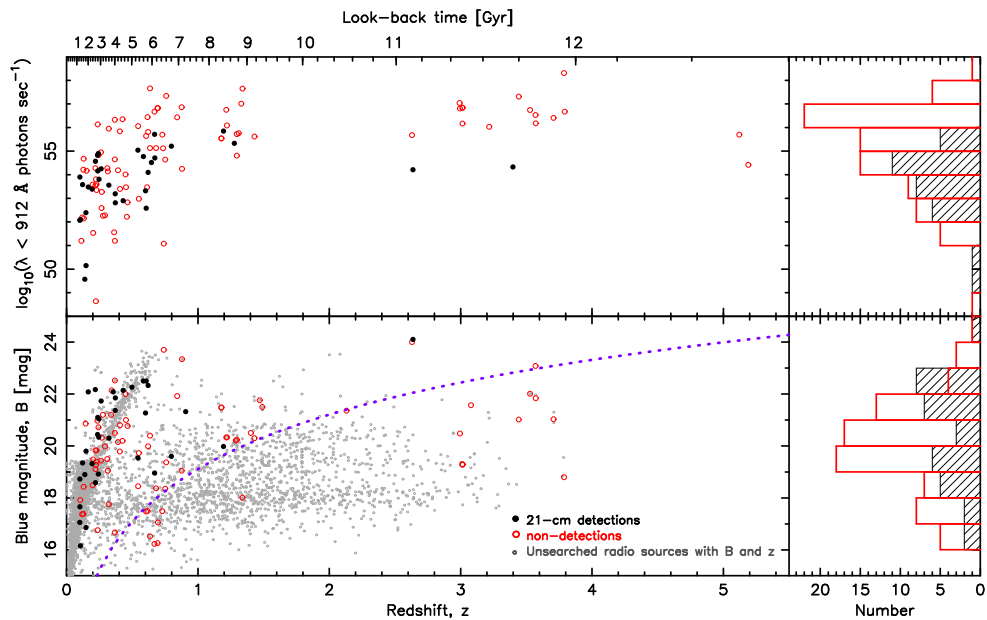

*Figure 4: Top: The ionising photon rate versus the redshift for the z ≥ 0.1 galaxies and quasars searched for HI 21cm absorption (Curran & Whiting 2012). Filled symbols/histogram show the detections and unfilled symbols/histogram the non-detections. Bottom: The corresponding optical blue magnitudes (updated from Curran et al. 2013). The dotted line shows the B magnitude corresponding to the critical UV luminosity identified by Curran & Whiting 2010. In objects below this line, all the gas in the galaxy is expected to be ionised rather than neutral (assuming a spectral slope of α =−1.5).*

## 2.3 Upcoming surveys with SKA pathfinders and precursors

Most of the work in the next few years will be focused on exploiting the large field of view of some of the SKA pathfinder and precursor telescopes, along with the uniform frequency coverage enabled by the radio-quiet SKA sites. This will make it possible to carry out blind searches that can probe the HI distribution in radio sources in a more systematic way. At this stage, the sensitivity of the pathfinders (at least for searches up to redshift z~0.2) will only be comparable to that of existing telescopes, though the available redshift range will be larger. These pathfinder observations will only be able to use relatively bright radio sources, telling us about gas in the central region of radio AGN with continuum flux > 50mJy (where we will be able to probe optical depth down to a few %).

Automated search techniques for HI absorption in blind surveys have already been developed (Allison et al. 2012, 2014; Koribalski 2012), along with techniques for stacking (Gereb et al. 2014a) and profile characterisation (Gereb et al. 2014b). The pathfinder and precursor surveys will allow us to expand the use of stacking techniques and derive general properties of group of objects (e.g. young/old/restarted, compact/extended and exploring trends with redshift, optical properties etc.). Furthermore, some of the new surveys will give us a first census of what to expect at higher redshift. The possibility of extending the search to redshift z~1 using ASKAP (FLASH survey, PI E. Sadler), with improved sensitivity and bandwidth compared to previous searches (with GMRT and WSRT) will be a first step in the study of evolution of HI content and its properties.

## 3. **Intervening absorption**

The SKA will be a uniquely powerful instrument for studying neutral gas in the distant universe through HI surveys for *intervening absorbers,* which will provide large samples of HI-selected galaxies spanning a huge range in redshift. Figure 5 shows the profile of intervening HI 21cm absorption detected along the line of sight to the bright continuum source PKS 1830-211 with the WSRT (Chengalur et al. 1999), and more recently with the first three ASKAP antennas.

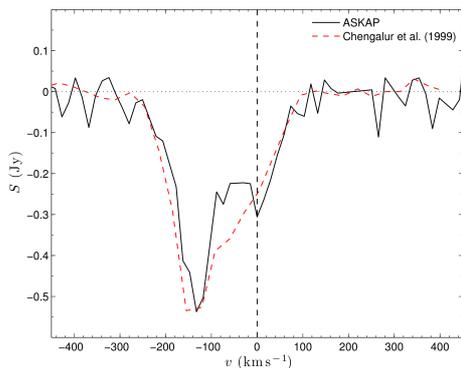

*Figure 5: An early ASKAP spectrum of intervening HI 21cm absorption at z = 0.89 along the line of sight to the strong radio source PKS 1830-211, a gravitationally-lensed QSO at z = 2.51. The line profile from the original detection by Chengalur et al. (1999) is shown in red for comparison.*

Almost all the currently-known intervening HI 21cm absorption-line systems were found through pointed observations of pre-selected targets, but SKA1 (and some of its precursor and pathfinder telescopes) will be able to carry out genuinely blind HI 21cm absorption surveys of large areas of sky, detecting many thousands of HI absorption systems and providing new insights into the role of neutral hydrogen in galaxy evolution across cosmic time.

### 3.1 Motivation and previous work

Cold neutral gas is central to understanding how galaxies and their star-formation rates evolve over cosmic time, yet we currently lack a clear picture of how much neutral gas is present in the Universe at different epochs or how this gas is distributed in dark halos of different mass. Almost all our current measurements of HI in the distant universe come from optical observations of Lyman α (Ly α) absorption by gas clouds along the line of sight to distant quasars (e.g. Wolfe et al. 1986; Lanzetta et al. 1991; Wolfe et al. 1995; Storrie-Lombardi, Irwin & McMahon 1996; Peroux et al. 2005; Wolfe, Gawiser & Prochaska 2005; Rao et al. (2006),

Prochaska & Wolfe 2009; Noterdaeme et al. 2012). With the exception of the study by Rao et al. (2006), these observations have mainly been carried out at redshift z > 2 where the UV Ly α line is redshifted to optical wavelengths and can be observed by ground-based telescopes. As a result, we currently have more accurate estimates of the cosmic mass density of HI at z > 2 than at lower redshifts (see Figure 6).

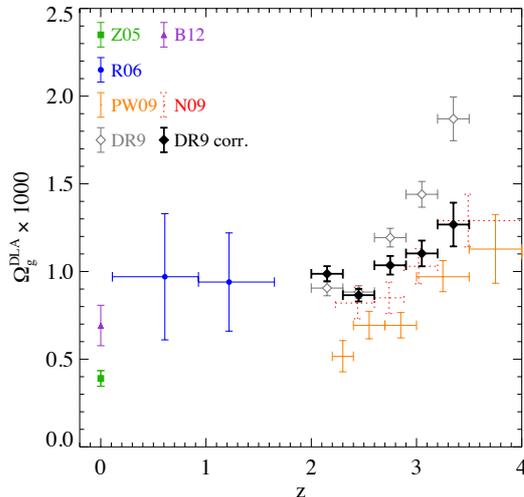

*Figure 6: Cosmological mass density of neutral gas in DLAs as a function of redshift, from Noterdaeme et al. (2012). Data points are coded as follows: Z05: Zwaan et al. (2005), B12: Braun (2012), R06: Rao et al. (2006), PW09: Prochaska & Wolfe (2009), DR9: Noterdaeme et al. (2012).*

Studies of the 21cm HI line in absorption along the line of sight to distant radio galaxies and quasars can provide a direct tracer of both the distribution of neutral hydrogen in distant galaxies and the physical conditions in the neutral gas (Kanekar & Chengalur 2003). Radio measurements of HI 21cm absorption offer some advantages over the existing optical Ly α surveys. There is no redshift limit for line detection, other than that set by the frequency coverage of the telescope and the availability of suitable bright background continuum sources. Radio observations are unaffected by extinction and can probe objects with a high dust content, in contrast to optical surveys for damped Ly α absorbers (e.g. Ellison et al. 2001). In addition, optical/IR identification and follow-up of HI-detected galaxies is likely to be easier if the background source is an optically-faint radio galaxy rather than a bright quasar (which generally overwhelms the light of a fainter intervening galaxy). On the other hand, because a 21cm absorption survey provides a measurement of HI optical depth rather than HI column density as discussed below (see also Braun 2012), interpretation of the data also presents some novel challenges.

Kanekar & Briggs (2004) laid out a detailed science case for HI 21cm absorption studies with the SKA. They suggested that blind SKA 21-cm surveys will yield large, unbiased absorber samples, tracing the evolution of normal galaxies and active galactic nuclei from z > 6 to the local universe. Such surveys will make it possible to directly measure the physical size and mass of typical galaxies as a function of redshift, and will provide new tests of hierarchical models of structure formation.

The Kanekar & Briggs (2004) paper remains an essential reference document for SKA studies, and our focus here is on developments over the past decade. There are three areas in particular where rapid progress has been made:

(1) Our understanding of **neutral molecular gas in distant galaxies** has advanced in recent years through observations with new mm and sub-mm interferometers. This opens up new opportunities to combine information about both the atomic and molecular phases of neutral gas in distant galaxies, providing a much more complete picture of their gas content as well as more rigorous constraints on theoretical models. A methodology for large "blind" mm-wave surveys for molecular absorption has been laid out, and a survey carried out with the GBT (Kanekar et al. 2014a; see also Curran et al. 2005). Such absorption surveys with ALMA and EVLA are likely to provide an important complementary view of the neutral ISM in high-z galaxies to that provided by HI 21cm absorption studies.

In a targeted study, Tacconi et al. (2010) found that the molecular gas content of spiral galaxies (as measured from the CO emission line) was three to ten times higher at redshift z=1-2 than it is today, and molecular gas has now been detected in QSO host galaxies out to z > 6 (e.g. Wang et al. 2011). Although these studies focused on bright objects, rapid progress in understanding the molecular gas properties of distant galaxies seems certain to continue as ALMA comes into full operation in the near future. There is already evidence that the overall molecular gas content of galaxies evolves more rapidly with redshift than their neutral hydrogen content over the redshift range $0 < z < 2$ (Lagos et al. 2011; see Figure 7), but as yet we have almost no observational constraints on the relative amounts of atomic and molecular gas in individual galaxies except in the very local universe.

(2) New HI absorption measurements of radio-loud QSOs have advanced our understanding of the **typical spin temperature and filling factor** in intervening HI 21cm absorption systems, particularly at z > 2 (Kanekar et al. 2014b and references therein). In particular, large GMRT and GBT studies of HI-21cm absorption in MgII-selected samples at z<1.7 (e.g. Gupta et al. 2007; Kanekar et al. 2009; Gupta et al. 2012) have significantly increased the number of known HI 21cm absorbers in the redshift desert, and provide a new route to finding DLAs at z<1.7, where the Lyman-alpha line is not observable from the ground. Around 200 QSO DLAs have now been studied in detail with optical telescopes, allowing the metallicity, elemental abundances and kinematics to be measured. The results of these studies imply that the number of absorbers with high spin temperature ($T_s > 1000$ K) increases with redshift (though Curran & Webb (2006) argue that part of this difference may be due to changes in the covering factor *f*). An important new result is the spin temperature - metallicity anti-correlation found in DLAs (Kanekar & Chengalur 2001; Kanekar et al. 2009). This anti-correlation is important for SKA1 searches at high redshift because it implies that galaxies with low metallicities may have high spin temperatures, and so may be more difficult to detect in HI 21cm absorption.

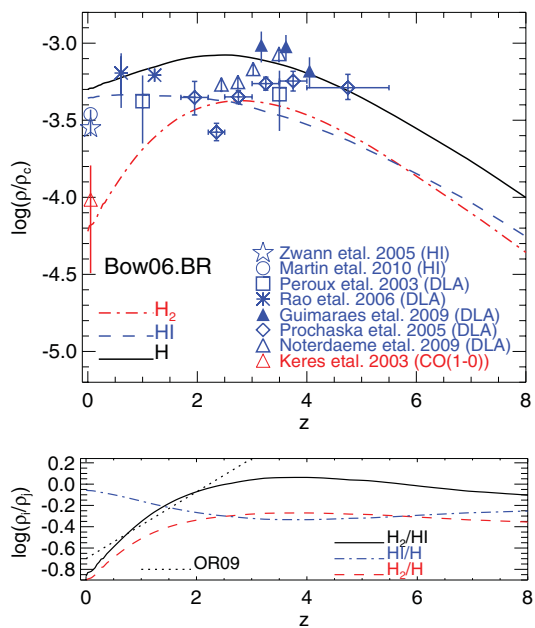

*Figure 7: (top) Global density of all forms of neutral hydrogen (solid line), atomic HI (dashed) and molecular $H_2$ (dot–dashed) in units of the critical density at z = 0, as a function of redshift; (bottom) evolution of the global $H_2$/H I (both plots from Lagos et al. 2011).*

(3) There have also been significant advances in **modelling the evolution of cold atomic and molecular gas** in distant galaxies, using both semi-analytic and hydrodynamical approaches. A detailed discussion of these models is presented in the chapter on *Neutral Hydrogen and Galaxy Evolution* (Blyth et al., this volume).

### 3.2 Forthcoming surveys with SKA pathfinders

As with the associated HI absorption case outlined in section 2.3, SKA pathfinder telescopes will carry out the first large blind imaging radio surveys for intervening HI 21cm absorption. ASKAP, with its 30 deg$^2$ field of view, and frequency coverage down to 700 MHz (i.e. redshift $z \sim 1$ for the HI line), will open up a large area of new parameter space for HI 21cm absorption studies, and Apertif and MEERKAT will also do important work. The ASKAP FLASH survey will cover the whole southern sky, searching more than 150,000 sightlines to bright (> 50 mJy) continuum sources for intervening HI absorbers in the redshift range $0.5 < z < 1$. The 5-sigma detection limit in HI optical depth for the planned integration time of 2 hours/field is expected to range from $\tau < 0.01$ for the brightest (> 1 Jy) background continuum sources to $\tau \sim 0.3$ for the faintest (50 mJy) sources. Thus the FLASH survey will mainly target the rare, high-column density absorbers expected to occur on < 1% of sightlines (Braun 2012). Deeper integrations with e.g. Apertif and MEERKAT (PI N. Gupta) will cover smaller areas of sky but probe to deeper HI optical-depth limits.

An important feature of all these pathfinder surveys is that unlike previous HI 21cm absorption studies, they will use all the strong continuum sources in the field (both radio galaxies and QSOs) to search for absorption lines. The angular size of these background sources will depend on several factors, including the source type and redshift, and many of them are likely to have resolved structures on sizes smaller than the angular resolution of the pathfinder telescopes (at 843 MHz, roughly 10% of continuum sources brighter than 50 mJy are expected to be QSOs; Jackson et al. 1998). Braun (2012) notes that the highest column-density HI gas in galaxy disks is likely to reside in discrete clumps with a characteristic scale size of ~100 pc (angular size 15 mas at $z \sim 1$), which will have a low filling factor if observed at arcsec resolution against a background radio galaxy several kpc in extent. As well as providing a large 'HI-selected galaxy sample' out to $z \sim 1$, high-resolution follow-up studies of HI 21cm absorption systems detected in these first blind surveys will provide important pathfinder results to guide and refine the design of future large HI absorption surveys with SKA1.

## 4. Expected science outcomes from SKA Phase 1

### 4.1 Planned observations

Table 1 outlines a set of three planned surveys for HI 21cm absorption with SKA1. These surveys are complementary to the strawman HI emission surveys proposed by Blyth et al. (this volume) but the use of absorption lines allows us to span a far larger redshift range than will be possible with the SKA1 emission-line studies (or even the range covered by HI emission studies with SKA2, which will reach to $z \sim 2$). Survey 1 is expected to take 1-2 months to complete at 30 km s$^{-1}$ spectral resolution, and will extend the pathfinder studies out to higher redshift by providing a detailed inventory of both intervening and associated HI 21cm absorption for galaxies in the redshift range 1 < z < 3. This will provide exciting new scientific results in its own right, but can also be used to refine the observing strategy for a larger-area survey (Survey 2 in Table 1) which would take ~1 year to complete and provide breakthrough science for understanding both galaxy feedback processes at high redshift and the co-evolution of cold gas and star formation across cosmic time. The third survey, with SKA1-LOW, will open new discovery space at z > 3, where the volume density of powerful radio galaxies is currently poorly known. This survey will be able to identify high-redshift radio galaxies and quasars through the 21cm line, with no requirement for optical pre-selection, and will connect the lower-redshift HI results to future studies of the HI '21cm forest' at z > 7 with SKA1 and SKA2 (e.g. Furlanetto & Briggs 2005; Ciardi et al. this volume).

*Figure 8: from Lagos et al. 2011 – needs a caption!*

| Survey | SKA 1 | Redshift range | Receiver | Spatial res. | Sky area | Spectral-line rms | Optical depth $\tau$ | Expected detections |
|---|---|---|---|---|---|---|---|---|
| 1. Inventory of HI in distant galaxies | SUR or MID | To z=3 | Band 1 | ~1 arcsec* | 1,000 deg$^2$ | <0.1 mJy | <0.01 (10mJy source) | ~5000 associated, several hundred intervening |
| 2. (i) Cold outflows (assoc.), (ii) Evolution of HI in galaxies (interven.) | SUR or MID | To z=3 | Band 1 | ~1 arcsec* | 10,000 deg$^2$ | <0.1 mJy | 0.001 to 0.005 (20mJy source) | A few hundred outflows, several thousand intervening absorbers |
| 3. HI at very high redshift | LOW | 3 < z < 8 | 220 MHz band | ~5 arcsec | >1,000 deg$^2$ | <0.5 mJy | <0.05 (10mJy source) | Unknown, new discovery space. |

\* *not possible with SKA1-SUR in band 1 in the baseline design*

*Table 1: Specifications of three proposed HI 21cm absorption surveys with SKA1. The listed spectral-line rms is indicative of the noise levels required to reach the science goals outlined in this paper*

**4.2 Associated absorption**

The main breakthroughs expected (for associated HI absorption) from SKA1 will be:

1) the use of absorption to trace the presence and properties of HI in active galaxies out to z~3 and beyond, as a function of galaxy properties and redshift;

2) tracing the presence of neutral gas outflows in sources below 100 mJy in flux, therefore increasing dramatically the numbers of potential candidates;

3) tracing the evolution of outflows (occurrence and characteristics) as a function of redshift, providing crucial information for this important ingredient in the feedback scenario;

4) making use of serendipitous detections of HI absorption to derive new redshifts (also for weak radio sources, i.e. not dominated by a radio AGN).

For studies of associated absorption with SKA1, we will be using Bands 1 and 2. This allows us to observe from redshift 0 (1421MHz, the HI 21cm rest frequency) to 350 MHz corresponding to redshift z~3 with MID and SUR. We will also use SKA1-LOW to probe associated absorption at z > 3, noting that associated systems are likely to have higher metallicity than intervening systems at these high redshifts (since they are close to a bright AGN) so cold gas may be easier to detect.

The success of searches for HI absorption will depend on the available sensitivity (see below). This will be combined with high spatial resolution observations because, as discussed above, pinpointing the location of the absorption is key for the interpretation. We will need at least arcsec/sub-arcsec resolution for this work, but possibly VLBI-like capabilities. This appears to be not well covered by the capabilities described in the Level 0 requirements (Dewdney et al. 2013; see also the chapter on *Very Long Baseline Interferometry with the SKA* by Paragi et al. this volume). Bandpass stability is also an extremely important parameter to allow the detection, also in weak radio sources, of shallow components and signatures of outflows (the stability of the bandpass is required to a level of 1 in $10^5$).

Present and near-future surveys can provide an inventory of HI 21cm absorption down to ~50 mJy for a typical optical depth (τ) of a few %. The big step forward provided by SKA1 is the possibility of extending the search and characterisation of the HI to higher redshifts (z > 1).

To make a substantial step forward and explore the presence of HI against radio sources down to the 10 mJy level, we will need to reach noise levels (per channel) < 100 μJy. Given the relatively high density of sources at this low flux level (around 15 per $deg^2$) it would not be necessary to cover a huge area: an area of roughly 1000 $deg^2$ would give ~15,000 sources, and if the detection rate holds to 30% (e.g. Gereb et al. 2014b) we will have ~5000 absorption systems to study in different redshift bins and to divide in different groups. The stacking will be done using weaker sources in the field as well.

Covering even larger areas would have the advantage that, for a large number of objects, we will be able to use HI 21cm absorption to derive the redshift of the host galaxy. For example, in the 3π sr survey from Fig. 5 and 7 of the Level 0 document, sources of ~30mJy can be searched for absorption down to ~ 2 % optical depth. An important goal is to trace the presence of outflows of cold gas as function of the radio power (i.e. exploring the low flux/radio power range). To trace these outflows with τ = 0.001-0.005 (typical of the known cases, Morganti et al. 2005) in sources of 100 mJy, we will need to reach ~30 – 170 μJy. Reaching these sensitivities will substantially increase the number of candidate sources that can be searched, allowing us to derive the statistics of the occurrence of cold gas outflows as a function of flux and of redshift.

To reach this level, we will need to go at least a factor of two deeper than the proposed SKA1 MID/survey 3π sr (Figs 5 and 7 in the Level 0 document) and cover at least 10,000 $deg^2$ of sky (from source counts we expect about 2 sources per $deg^2$ above 100 mJy) to allow enough sources to be explored. If the detection rate of outflows is about 10% of the overall HI 21cm absorption *detections* (e.g. Gereb et al. 2014b), this would give ~ 600 potential outflows with which to derive the statistics of their characteristics and evolution in different redshift bins. For this particular science case, the full velocity resolution of 30 km $s^{-1}$ will not strictly be required, and therefore the observing time can also be slightly reduced. A shallow all-sky survey will not provide, for this topic, a comparably interesting step forward. The expected spatial resolution of SKA1 is higher than 1arcsec only up to 1 GHz. This is likely to limit our ability to pinpoint the location of the absorbing HI gas and outflows.

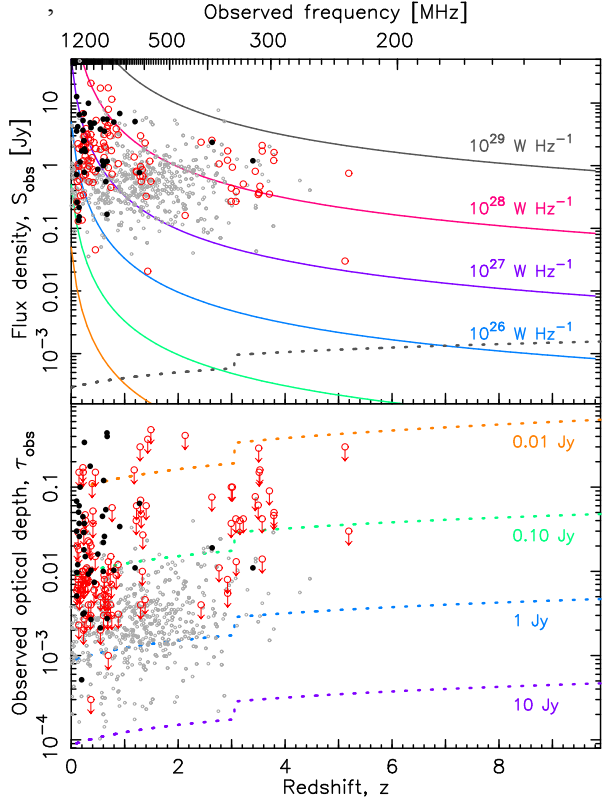

*Figure 8: (Top): Flux densities of sources at z > 0.1 which have so far been searched for HI absorption. As in Fig. 4, filled symbols show detections and open symbols non-detections. The small grey circles show Parkes Quarter-Jansky (Jackson et al. 2002) sources with known redshifts. Current z > 0.1 searches range from $10^{24}$ - $10^{29}$ W Hz$^{-1}$ (Curran et al. 2008), and the dashed line shows the rms noise level of SKA1–MID and SKA1–Low in a 1 km s$^{-1}$ channel in a 1 hour integration. (Bottom): The corresponding optical depths and 3σ limits. Dashed lines show the sensitivity of SKA1 for the corresponding flux density at 1 km s$^{-1}$ resolution after 1 hour.*

Figure 8 (top) shows the flux densities of sources so far searched for associated HI absorption at redshift z > 0.1. There are two decades of unexplored space between these observations and the SKA1 sensitivity. SKA1–MID and –LOW will cover the entire frequency range from 50 to 1420 MHz in just a few tunings, means that there is no need to preselect targets using an optical redshift. This opens up unexplored territory at high redshift, possibly uncovering a population of obscured gas-rich radio galaxies and QSOs that are hidden at optical wavelengths.

**Synergy with other facilities**. The SKA1 studies will provide extremely interesting targets for follow up with the optical and mm facilities that also available in the southern hemisphere. In particular, the study of the HI outflows will be relevant and complementary to the molecular studies. The wider field of view of HI observations will allow blind searches and provide targets for ALMA. Future large-area spectroscopic surveys in the southern hemisphere will also be an important complement to the HI data.

### 4.3 Intervening absorption

HI 21cm absorption-line surveys with SKA1 will provide an **HI-selected galaxy sample** extending out to z ~3 with SKA1-MID/SUR, and potentially at z >> 3 with SKA1-LOW. Follow-up observations with ALMA will allow us to link HI and CO along the same sightline for galaxies over a wide range in redshift, and follow-up optical/IR studies will also be possible for many objects out to at least z ~ 1.5 to 2. At redshifts above z ~ 2, there will be a significant overlap with radio-loud QSOs for which optical spectra of the Ly α line are already available.

An SKA1 HI 21cm absorption survey will also be able to find heavily reddened and dust-obscured QSOs that have been missed by optical surveys.

At z > 1, SKA1 HI absorption studies will open up a new parameter space not explored by the pathfinder surveys. SKA1 will provide detailed information on the HI properties of individual galaxies at lookback times greater than 8 Gyr. A wide field of view is an advantage for these studies, because the surface density of sufficiently distant background radio sources is low, and decreases out to higher redshift (e.g. the expected surface density of continuum sources brighter than 30 mJy is roughly 2.5 per deg$^2$ for sources with z > 1 and 1 per deg$^2$ for sources at z > 3; see Kanekar & Briggs 2004 and Wilman et al. 2008). Figure 9 shows the expected distribution of HI column densities at different redshifts, as discussed by Braun (2012).

A large-area survey of ~10,000 deg$^2$, as outlined in Table 1, would allow 5-$\sigma$ detection of HI absorption lines with an optical depth of $\tau \sim 0.015$ against background sources stronger than about 30 mJy, and of stronger lines against weaker sources (e.g. $\tau \sim 0.05$ lines against sources stronger than ~10 mJy and $\tau \sim 0.1$ lines against sources stronger than ~ 2-3 mJy). The HI column densities probed by these lines would depend on the spin temperature and filling factor of the neutral gas, but such a survey would probe at least 50,000 sightlines to background radio sources at z > 1, and might be expected to detect several thousand intervening HI absorption systems out to redshift z = 3. Longer integration times could also be built up over smaller regions of sky to push to lower optical depth and provide better coverage of absorbers linked to warmer gas with spin temperatures above 200-300 K, providing a unique probe of the warm neutral medium in high-redshift DLA systems.

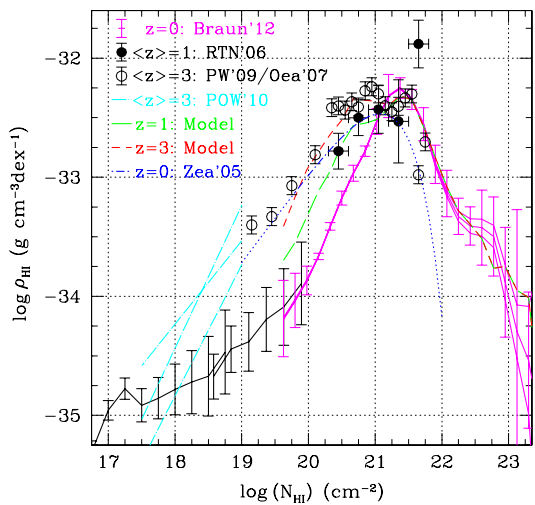

*Figure 9: HI mass density function for intervening HI absorption systems over the redshift range 0 < z < 3, from Braun (2012). Filled circles show the ⟨z⟩ = 1 QSO absorption line data of Rao et al. (2006), and open circles the ⟨z⟩ = 3 QSO absorption line data of Prochaska & Wolfe (2009) and O'Meara et al. (2007). The relation at z=0 is derived from 21cm HI emission surveys*

**Synergy with other facilities**. HI 21cm absorption surveys will provide data for an exciting range of follow-up programs with other facilities. Current optical/UV DLA samples are dominated by systems at redshift z > 1.7, with over 10,000 DLAs known at z > 2 and only about 50 at z < 1.7 (e.g. Noterdaeme et al. 2009; Lee et al. 2013). Redshifted HI-21cm absorption surveys with SKA1-SUR and –MID will therefore have a huge impact on studies of neutral gas in galaxies at z < 1.7, both through blind surveys and by targeting QSOs with known MgII absorbers. In particular, blind HI 21cm absorption surveys with SKA1 will provide large and unbiased samples of HI-selected galaxies across a wide range in redshift. We can then probe the nature of "typical" galaxies as a function of redshift, via follow-up optical/IR imaging with large ground-based telescopes and JWST, and spectroscopy using optical/near-IR integral field spectrographs to study the Ly$\alpha$ and H$\alpha$ lines in emission, and CO studies with ALMA.

SKA1 will also have an impact at z > 1.8, where large ground-based studies of QSO absorption have already been carried out with optical telescopes. At least 10% of the powerful radio sources used to search for HI 21cm absorption with SKA1 are expected to be bright optical QSOs, and combining optical and radio measurements for these objects will provide new insights into the physical properties of the interstellar medium at high redshift through measurements of the spin temperature and metallicity for large samples of HI absorption systems (Kanekar et al. 2014).

The other place where HI 21cm absorption surveys will have an important impact is at z > 5, where it becomes difficult to identify DLAs at optical/IR wavelengths because the Lyman-alpha forest becomes extremely dense. The limiting factor here is the low surface density of bright radio continuum sources at z > 5, but HI absorption surveys with SKA1-LOW should yield at least a modest number of detections which can be followed up at other wavelengths (though we note that ALMA CO emission studies may be challenging at z > 5, since the CO-to-H2 conversion factor is likely to be very high in small, low-metallicity galaxies at high redshift).

Over the full SKA 1 redshift range, blind HI 21cm absorption surveys will also find dusty absorbers that could be followed up to search for OH/CO/HCO+ absorption. Such observations would provide tests of fundamental constant evolution, as well as allowing us to study molecular gas in normal galaxies over a wide range in redshift. In addition, optical measurements of molecular hydrogen (H$_2$) absorption along the sightline to 21cm HI absorption systems will allow a direct comparison of the atomic and molecular phases of the neutral ISM at high redshift (Ledoux et al. 2003; Hirashita et al. 2003). ALMA observations of CO emission within distant galaxies identified through intervening HI absorption will also allow us to study correlations between the 21cm optical depth or spin temperature, H$_2$ absorption and CO emission within the same objects, making it possible to link the physical conditions derived from HI absorption-line studies (with SKA and optical telescopes) and molecular emission-line studies with ALMA.

## 5. Expected outcomes from early science (50% SKA1)

The main breakthrough enabled by SKA1 is the opening-up of new parameter space for studies of HI absorption at redshift 1 < z < 3 (350-700 MHz), a redshift range not covered by the SKA pathfinder and precursor telescopes. Existing telescopes have probed some regions of this redshift space, but bandwidth limitations and the effects of radio-frequency interference (RFI) have strongly limited their redshift coverage. The radio-quiet site of SKA1 offers an important advantage for work in this frequency range. Although a 50% SKA1 instrument will be less sensitive (in terms of Ae/Tsys) than existing large telescopes like GMRT and GBT, it will still open up an exciting new parameter space for HI 21cm absorption surveys in the distant universe. For example, a 50% SKA1 instrument would have a survey speed roughly 25 times faster than the GMRT for blind HI absorption surveys at 350-700 MHz. HI absorption studies are an excellent early science project because the strongest lines have high optical depth and can be detected in fairly short integrations. Since SKA1 opens up a new frequency range below 700 MHz, new science can be done even with much less than 50% of the final collecting area.

## 6. Anticipate the science outcomes of SKA 2

For intervening HI 21cm absorption, a key focus for SKA2 will be the evolution of cold gas in galaxies across cosmic time, including direct measurements of the physical size and mass of typical galaxies as a function of redshift out to $z > 6$, as discussed in detail by Kanekar & Briggs (2004). Beyond $z \sim 6$, HI 21cm absorption measurements potentially provide a unique measurement of the HI density power spectrum on small scales (Ciardi et al., this volume). For associated HI absorption, SKA2 will provide two main steps forward. The enhanced sensitivity at low frequencies will allow us to trace the evolution of radio galaxies (not only the powerful one) and, in particular, the inflow and outflow of cold gas with their host galaxies, over almost the whole of cosmic time. The higher spatial resolution offered by SKA2 will also be a major advantage for the interpretation of associated HI detections and their links to the nuclear activity and the evolution of the host galaxy.